\documentclass[manuscript]{aastex}

\def\kms  {km~s$^{-1}$}
\def\masy {mas~y$^{-1}$}

\def\etal {et al.~}

\def\apj  {$Astrophys. \ J.$}
\def\apjs  {$Astrophys. \ J. \ Supp.$}
\def\aap  {$Astron. \ Astrophys.$}
\def\apjl  {$Astrophys. \ J. \ Lett.$}
\def\mnras  {$Mon. \ Not. \ R. \ Astron. \ Soc.$}
\def\aj  {$Astron. \ J.$}

\slugcomment{Version: 8 December 2005}

\shorttitle{Distance to W3OH} \shortauthors{Xu et al.}

\begin{document}

\title{The Distance to the Perseus Spiral Arm in the Milky Way}

\author{Y. Xu\altaffilmark{1,2,3}, M. J. Reid\altaffilmark{2}, X. W. Zheng\altaffilmark{1,2},
K. M. Menten\altaffilmark{4}}

\altaffiltext{1}{Department of Astronomy, Nanjing University
Nanjing 210093, China, yxu@shao.ac.cn;\ xuye@mpifr-bonn.mpg.de}
\altaffiltext{2}{Harvard-Smithsonian Center for Astrophysics, 60
Garden Street, Cambridge, MA 02138, USA, reid@cfa.harvard.edu}
\altaffiltext{3}{Shanghai Astronomical Observatory Chinese Academy
of Sciences, Shanghai 20030, China}
\altaffiltext{4}{Max-Planck-Institut f$\ddot{u}$r Radioastronomie,
Auf dem H$\ddot{u}$gel 69, 53121 Bonn, Germany}

\begin{bf}
{We have measured the distance to the massive star-forming region W3OH
in the Perseus spiral arm of the Milky Way to be 1.95 $\pm$ 0.04~kilo-parsecs
($5.86\times10^{16}$~km).  This distance was determined by triangulation,
with the Earth's orbit as one segment of a triangle, using the Very Long Baseline
Array.  This resolves a long-standing problem of a factor of two discrepancy
between different techniques to determine distances.  The reason for the
discrepancy is that this portion of the Perseus arm has anomalous motions.
The orientation of the anomalous motion agrees with spiral density-wave theory, but
the magnitude is somewhat larger than most models predict.
}
\end{bf}




Massive stars and their associated bright regions of ionized
hydrogen trace the spiral arms of galaxies.  However, for our galaxy,
the Milky Way, our view from the interior makes it difficult to
determine its spiral structure.  In principle, one can
construct a simple model of the rotation speed of stars and gas
as a function of distance from the center of the Milky Way.
Then, if one measures the line-of-sight component of the velocity
of a star or interstellar gas, one can determine its
distance by matching the observation with the model prediction
(i.e., a kinematic distance).  Knowing distances to star forming
regions, one can then locate them in 3-dimensions and
construct a ``plan view"--a view from above the plane--of the
Milky Way.  Unfortunately, many problems arise when constructing a
plan view of the Milky Way, including 1) difficulties in
determining an accurate rotation model (which requires the distance
and orbital speed of the Sun from the center of the Milky Way),
2) distance ambiguities in some portions of the Milky Way (where an observed
velocity can occur at two distances), and 3) departures
from circular rotation (as might be expected for spiral structure).
Progress has been made on the first two problems. For example, many
kinematic distance ambiguities can be resolved by interferometric
studies of hydrogen absorption in the radio, since distant sources
will show a greater velocity range for hydrogen absorption than near
sources {\it [1]}.
However, the third problem, non-circular motions, is fundamentally much harder to
address.

The Perseus arm is the nearest spiral arm outward from the Sun, as shown in
Fig.~1 {\it [2, 3]}.
There are many star-forming regions in the Perseus arm
for which distances have been estimated from the difference between the
observed and intrinsic luminosities of massive, young (O-type) stars.
Toward Galactic longitudes 132 -- 138$^{\circ}$, such luminosity
distance estimates are $\approx2.2$~kpc {\it [4]}. However, kinematic
distances for these regions are much greater.
Stars and gas in this region of the Perseus arm are observed to move with
line-of-sight velocities of about $-45$~\kms, relative to the Local Standard of
Rest (LSR) {\it [5]}.
Assuming that the Milky Way rotates at 220~\kms, independent of distance from
its center, and that the Sun is at a distance of 8.5~kpc from the center
{\it [6]}, the observed Perseus arm velocities occur at
distances of $>4$~kpc.  The discrepancy between distances based on
stellar luminosities and velocities has never been resolved.

\begin{figure}
\epsscale{0.6} \plotone{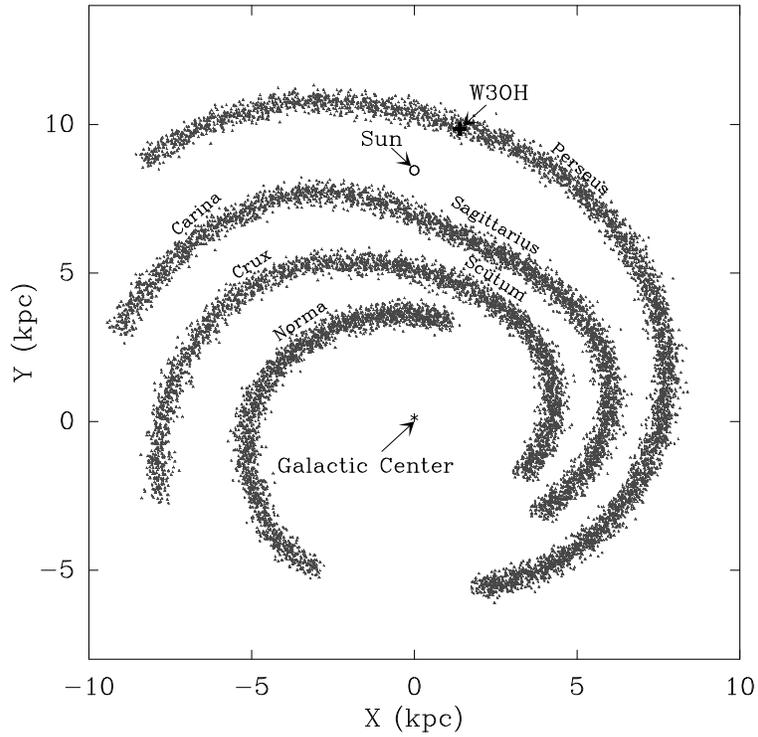}
\caption{
``Plan view'' of the Milky Way as seen from its north pole.
Estimated locations of spiral arms {\it [2, 3]} are indicated
by the large number of dots and labeled by a prominent constellation
onto which they are projected.
The locations of the Galactic Center, Sun, and W3OH are indicated.
 \label{fig1}
}
\end{figure}

The problem of the distance to the Perseus arm can be resolved by
determining an accurate distance to a massive star-forming region
in the arm.  The best and most reliable method for measuring
distance in astronomy is called a trigonometric parallax.
A trigonometric parallax is determined by observing
the change in position of star, relative to very distant objects
such as quasars, as the Earth moves in its orbit about the Sun.
The parallax is simply the maximum angular deviation of the apparent
position from its average position over a year.
The deviation in position of a source over a year is very small.
For example, the parallax for a source at a distance of 2~kpc,
or about one-quarter of the distance to the center of the Milky Way,
is only 0.5 milli-arcsecond (mas).

The distance, $D$, to a source is easily calculated from its parallax, $\pi$,
by triangulation: $D{\rm(kpc)} = 1 / \pi{\rm(mas)}$.  Thus, one needs a
measurement accuracy of 0.05 milli-arcseconds (mas) to achieve 10\% accuracy
for a source at 2~kpc distance, which would be sufficient to
resolve the Perseus arm discrepancy.  By comparison, trigonometric
parallaxes obtained by the Hipparcos satellite {\it [7]} typically have
uncertainties of only $\approx1$ mas, which is inadequate for our purposes.

An ideal candidate for a trigonometric parallax measurement is the
massive star forming region W3OH, located at Galactic longitude $134^\circ$
and near O-type star associations.  W3OH has strong methanol masers {\it [8]},
which can serve as bright, relatively stable, beacons for astrometric
observations at radio wavelengths.  In this paper, we describe observations
with the Very Long Baseline Array (VLBA), consisting of ten radio telescopes
spanning the Earth from Hawaii to New England to the Virgin Islands and
operated by the National Radio Astronomy Observatory {\it [9]},
which allowed us to achieve an extraordinarily accurate (0.01~mas)
parallax for W3OH and to resolve the long-standing problem of the
distance to the Perseus arm.  Our results also provide
valuable information to test spiral density-wave models of the Milky Way.

We observed W3OH and three compact extragalactic radio sources
for 8 hours on each of five epochs in order to measure
the position of W3OH (using its methanol masers as
astrometric targets) relative to extragalactic radio
sources.   The dates of observation were 2003 July 30 and October 21,
and 2004 January 30, April 23, and July 25.  These dates well sample
the peaks and nulls of the sinusoidal trigonometric parallax signature,
caused by observing the source at different positions in the Earth's
orbit about the Sun.  This sampling ensures that we can separate
the linear proper motion (caused by projections of Galactic
rotation, as well as the peculiar motion of W3OH and the Sun) from the
sinusoidal parallax effect.

We switched rapidly among W3OH and the background sources, repeating
the following pattern: W3OH, J0235+622, W3OH, J0231+628, W3OH,
J0230+621. Sources were changed every 40 s, typically achieving 30~s
of on-source data each time.  We used a methanol maser as the
phase-reference source, because it is considerably stronger than
the background sources and could be detected on individual interferometer
baselines with signal-to-noise ratios exceeding $100$ in
the available on-source time.

We discuss details of the calibration procedures in supporting online material.
After calibration, we made an image of a strong reference maser channel.
We show the first and last epoch images of the reference spectral channel
at a LSR velocity, $v_{\rm LSR}$, of $-44.2$~\kms\ in Fig.~2.
One can see that there
is little change in the masers over a year.  In Fig.~S1
we show an image at one epoch of each of the background radio
sources.  All three background sources are compact ($<0.5$~mas)
and dominated by a single component.

\begin{figure}
\epsscale{0.5} \plotone{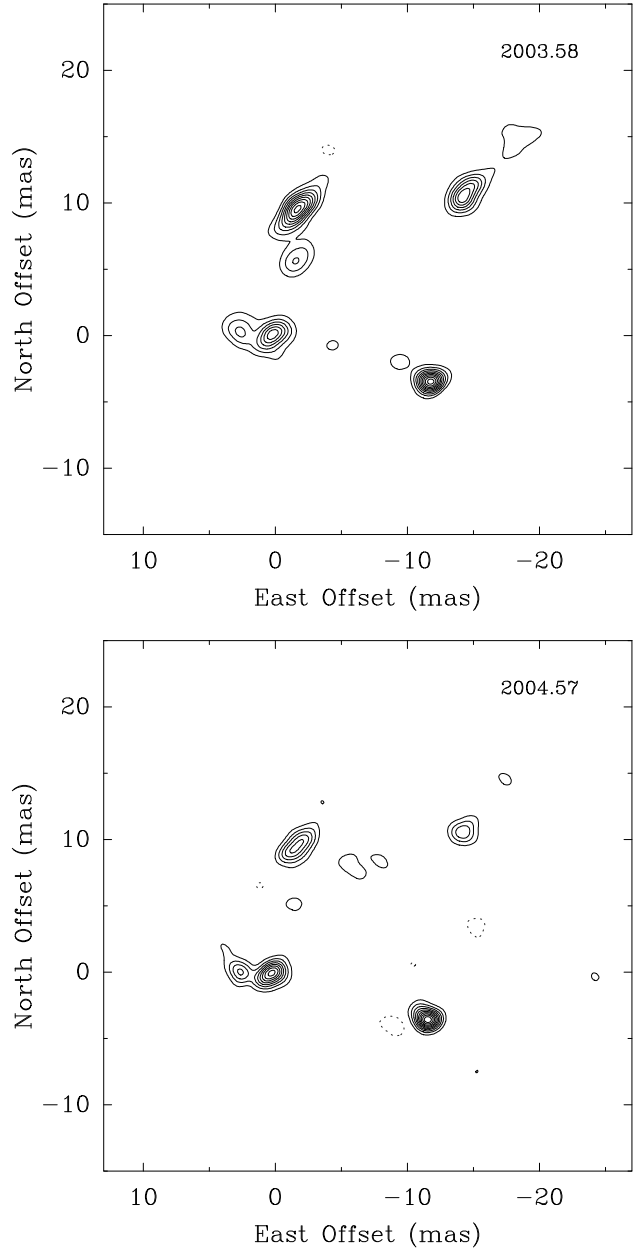} \caption{The first and
last epoch maps (at dates indicated in the upper right corner)
of 12 GHz methanol masers toward W3OH.  This map contains
emission from one spectral channel at $v_{\rm LSR}=-44.2$~\kms;
the maser reference spot at (0,0).  The
maser emission structures change little over the time range of our
parallax measurements.
Only the two brightest and most compact spots near (0,0) and
(-12,-3) mas were used for the parallax measurement.
The contour levels are integer multiples of 2 Jy beam$^{-1}$,
with negative values dashed and the zero contour suppressed.
The restoring beam size is 1.2~mas.   The origin of the maps is
$\alpha({\rm J2000})=02~27~03.8192$ and $\delta({\rm
J2000})=61~52~25.230$. \label{fig2}}
\end{figure}

In order to provide the data needed to measure the parallax and
proper motion (on the plane of the sky),
we fitted 2-dimensional Gaussian brightness distributions to the
nine brightest maser spots and the three background radio sources
for all five epochs. In Fig. 3 we plot the positions of one maser
spot in W3OH relative to the three background radio sources. The
change in position of a maser spot relative to a background radio source
was then modeled by the parallax sinusoid in each coordinate,
completely determined by one parameter (the parallax), and a
linear proper motion in each coordinate. Tables 1 and 2 list the
parameters of the fits.

An unweighted average parallax for the nine maser spots of W3OH measured
against each of the background sources yields parallax estimates of
$0.502 \pm 0.011$ mas using J0230+621,
$0.526 \pm 0.014$ mas using J0231+628, and
$0.515 \pm 0.015$ mas using J0235+622.
These results are consistent within their formal errors, and a weighted
average of these three parallaxes yields $0.512 \pm 0.007$ mas.

The parallaxes from the three calibrators increase slightly with
increasing maser--calibrator separation, suggesting that atmospheric
systematics may not have been entirely removed.
However, with only three calibrators it is difficult to assess the
significance of this effect.  In order to allow
for the possibility of some uncompensated atmospheric systematics,
we estimate a systematic component of the parallax uncertainty of 0.007~mas.
Thus, our final parallax estimate is $0.512 \pm 0.007$ ${\rm
(statistical)} \pm 0.007~{\rm (systematic)}$ mas.
The statistical and systematic uncertainties are independent
and combining them quadratically yields a parallax of $0.512 \pm 0.010$
and, hence, a distance of $1.95 \pm 0.04$ kpc.

Our trigonometric parallax estimate is consistent with a similar
measurement of $2.04 \pm 0.07$~kpc using H$_2$O masers associated
with the Turner-Welch (TW) object, a proto-stellar object projected 5~arcsec
from W3OH {\it [15]}.  These distances for W3OH in the Perseus spiral arm
conclusively resolve the long-standing discrepancy between its
kinematic distance of 4.3 kpc and a
luminosity distance of $\approx2.2$~kpc,
based on O-type stars nearby in the same spiral arm,
{\it [4]}.  The luminosity distance is consistent with the trigonometric
parallaxes, and W3OH must have a large kinematic anomaly.

\begin{figure}
\epsscale{0.6} \plotone{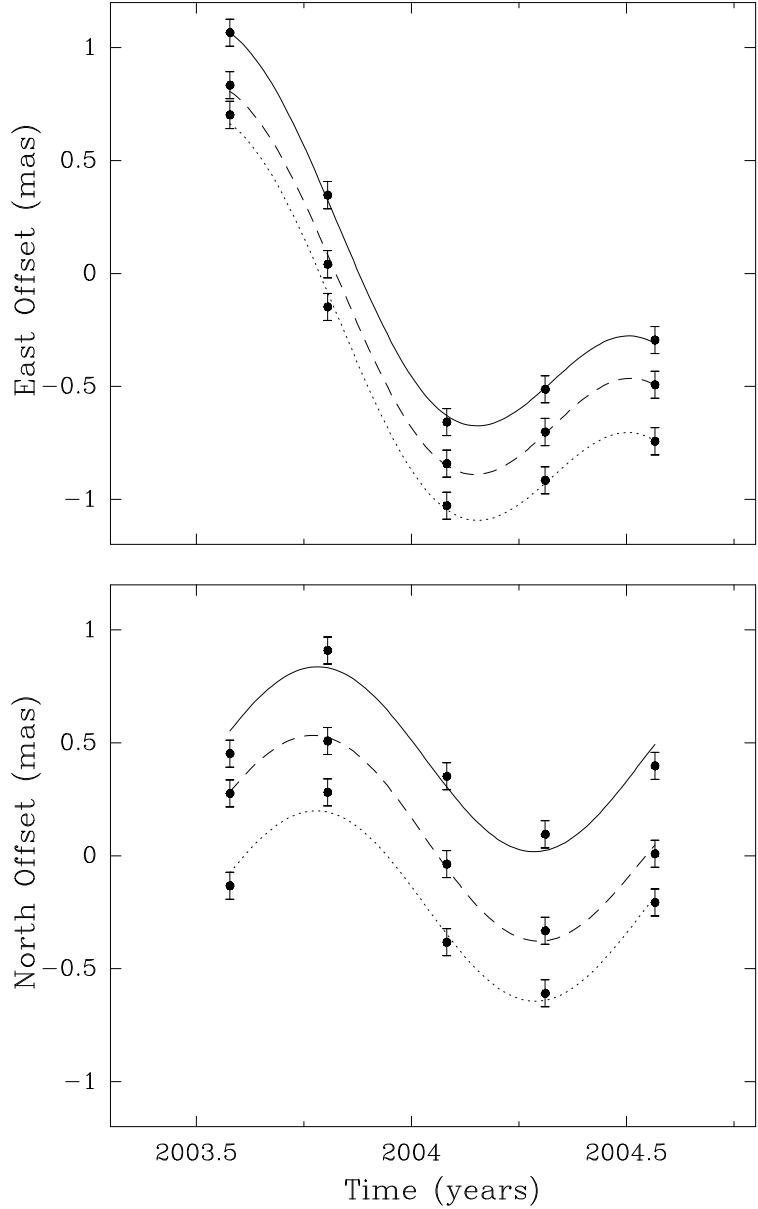}\caption{Position versus
time for the reference maser spot (i.e., the strong maser spot
near (0,0) in Fig.~2) at $v_{\rm LSR}= -44.2$ \kms\ relative
to three background radio sources.
The top and bottom panels show the eastward and northward offsets,
respectively.   The large difference in position
between W3OH and each background source has been removed and the
data for the different background sources have been offset for
clarity. In each panel, the top, middle, and bottom data are
for the background sources J0230+621, J0231+628, and J0235+622,
respectively. Also plotted are the best fitting models, specified
by five parameters: one for the parallax and two for the proper
motion in each coordinate.\label{fig3}}
\end{figure}

Table 2 presents the proper motion results. The individual proper
motions span a range of about 0.3 and 0.2~\masy\ (or 3 and 2 \kms)
in the eastward and northward directions.
This is comparable to the spread in hydroxyl (OH) maser proper
motions from the same region of W3OH {\it [10]}.  Thus, the dispersion in the proper
motions is likely to come both from small internal motions of the
maser spots of a few \kms\ and from measurement error of $\pm0.1$
\masy ($\pm1$ \kms). Because of the limited number of methanol
masers mapped and the small velocity spread expected for methanol
masers (based on the close correspondence with OH masers), we make
no attempt to fit an expanding model to the data.

The mean proper motion in the eastward and northward directions,
obtained from an unweighted average of all the data, is
$-1.204 \pm 0.02$ and $-0.147 \pm 0.01$ \masy, respectively, where
the uncertainties are standard errors of the mean.  Thus, the
uncertainty in the mean proper motion translates to an impressive
0.2 \kms\ at a distance of 1.95 kpc.  Allowing for a spread of
$\pm 3$ \kms\ for the internal motions of the masers, we adopt
a $1\sigma$ uncertainty of 1~\kms\ in each component of the
average of our motions.

In order to study the 3-dimensional motion of W3OH in the Galaxy, we
converted its radial and proper motion from the equatorial,
heliocentric, reference frame in which they are measured into a
Galactic reference frame.  A convenient Galactic frame is one
moving with a circular velocity about the center of the Galaxy
at the position of W3OH, i.e., a ``local standard of rest'' for
W3OH.  We followed published methods {\it [11]},
adopting the International Astronomical Union standard values
for the distance from the Sun to the Galactic Center of 8.5~kpc and for
the rotation speed of the LSR about the Galactic Center
of 220~\kms.  When removing the peculiar motion of the Sun relative to the LSR,
we adopted solar motion values determined by Hipparcos {\it [12]}.

The Galactocentric distance of W3OH is 9.95 kpc, and its rotation
velocity about the Galactic Center is $206 \pm 10$ \kms.
This velocity uncertainty is far above our measurement error
and is dominated by an uncertainty of about $\pm10$ \kms\ for
the rotation speed of the Galaxy.
However, the {\it difference} in the rotation
velocities of the Sun and W3OH is largely insensitive to the
value adopted for Galaxy's rotation speed.
Assuming a constant rotation speed in the Galaxy
between the 8.5 and 9.95~kpc (the Galactocentric radii of the Sun
and W3OH), the difference between the rotation speed of
W3OH and the Galaxy is $14 \pm 1$ \kms\,
in the sense that W3OH is orbiting slower than the Galaxy spins.
The motion of W3OH toward the Galactic Center is $17 \pm 1$ \kms.
The motion of W3OH toward the north Galactic pole is $-0.8 \pm 0.5$ \kms.
Combining the three components, the total peculiar motion of W3OH
is 22~\kms.  Essentially all of the peculiar motion is
in the plane of the Galaxy, as expected for a massive star-forming region.

Note that for an axially symmetric distribution of mass in the Galaxy,
rotational velocities cannot fall more rapidly with Galactocentric
radius, $r$, than $1/\sqrt{r}$.
The 14~\kms\ slower rotation of W3OH, compared to its local standard
of rest, would therefore require essentially no mass
between the Galactocentric radius of the Sun (8.5 kpc) and W3OH (9.95 kpc).
This seems unlikely and provides motivation for a non-axisymmetric
mass distribution, such as provided by spiral density-wave theory.

Spiral density waves might be able to account for some of
the peculiar motion of W3OH.
In order to maintain a spiral density wave, material flowing
into a trailing arm would be expected to acquire motion components
inward and counter to Galactic rotation {\it [13]} as we have observed.
However, for density contrasts between arm and interarm regions of
$\approx10$\%, peculiar velocity components of only
$\approx5$ \kms\ are expected.  This is smaller than we observe
for W3OH.  However, calculations suggest that
a spiral shock embedded in a background density wave
might lead to velocity jumps in molecular material of the
magnitude that we observe for W3OH {\it [14]}.

Gravitational forces from nearby forming stars, such as the
TW object, and massive molecular
clouds should contribute $\approx5$ \kms\ to the motion of W3OH.
Indeed, the TW object has been observed to move at
$\approx10$ \kms\ with respect to W3OH {\it [15]}.
Depending on the relative masses of these and
other stars and gas in the region, some of the 22~\kms\ peculiar
motion of W3OH might be explained by local gravitational effects.
Additionally, the giant molecular cloud material that existed
prior to the formation of the stars in the W3OH region could have
been accelerated by shocks associated with supernovae in the
region {\it [16-18]}. Overall, it remains to be seen if one can account
for the peculiar motion of W3OH in the context of the spiral
density-wave paradigm.

We have established that the VLBA can achieve a parallax accuracy
of 0.01~mas and proper motion accuracy of better than 1~\kms\ for
Galactic sources with only 5 observations spanning one year. With
this accuracy, the VLBA can be used to measure distances to 10 kpc
with better than 10\% accuracy, approximately a factor of 100 better
than the Hipparcos satellite.  Based on these results, we
believe that the VLBA, and ultimately the Japanese Very Long Baseline
Interferometric project VERA (VLBI Exploration of Radio Astrometry,
{\it [19]}), can map the spiral structure and full kinematics of
massive star forming regions in the Milky Way.

\begin{bf}{References and Notes}\end{bf}\\
1. V. L. Fish, M. J.  Reid, D. J. Wilner, E. Churchwell, \apj \ 587, 701 (2003)\\
2. Y. M. Georgelin, Y. P. Georgelin, \aap \ 49, 57 (1976)\\
3. J. H. Taylor, J. M. Cordes, \apj \ 411, 674 (1993)\\
4. R. M. Humphreys, \apjs \ 38, 309 (1978)\\
5. The LSR is a reference frame at the position of the Sun and
   moving in a circle about the center of the Milky Way; in practice this
   frame is determined from the average motion of large numbers of stars in
   the solar neighborhood.\\
6. F. J. Kerr, D. Lynden-Bell, \mnras \ 221, 1023 (1986)\\
7. M. A. C. Perryman, \etal \aap \ 323, L49 (1997)\\
8. K. M. Menten, M. J. Reid, J. M. Moran, T. L. Wilson, K. J. Johnston,
   W. Batrla, \apj \ 333, L83  (1988);
   L. Moscadelli, K. M. Menten, C. M. Walmsley, M. J. Reid, \apj \
   519, 244 (1999)\\
9. The National Radio Astronomy Observatory is operated by Associated
   Universities Inc., under cooperative agreement with the National Science
   Foundation.\\
10. near the origin in the maps of E. E. Bloemhof, M. J. Reid, J. M. Moran,
    \apj \ 397, 500 (1992)\\
11. D. R. H. Johnson, D. R. Soderblom, \aj \ 93, 864 (1987)\\
12. W.Dehnen, J. J. Binney, \mnras \ 298,387 (1998)\\
13. C. C. Lin, C. Yuan, F. H. Shu, \apj \ 155, 721 (1969)\\
14. W. W. Roberts, \apj \ 259, 283 (1972)\\
15. K. Hachisuka, \etal to appear in \apj (2006)\\
16. B. Dennison, G. A. Topasna, J. H. Simonetti, \apj \ 474, L31 (1997)\\
17. R. J. Reynolds, N. C. Sterling, L. M. Haffner, \apj \ 558, L101 (2001)\\
18. M. S. Oey, A. M. Watson, K. Kern, G. L. Walth, \aj \ 129, 393 (2005)\\
19. M. Honma, N. Kawaguchi, T. Sasao, in Proc. SPIE Vol.4015 Radio
    Telescope, ed H. R. Butcher, p624 - p631 (2001)\\
20. YX and XWZ thank the Smithsonian Institution for support through its
    visiting scientist program.  Research on the structure of the Milky Way
    at Nanjing University is supported by the National Science Foundation of
    China under grants 10133020 and 10373025.\\

\begin{deluxetable}{crrcccc}
\tablecolumns{7} \tablewidth{0pc} \tablecaption{W3OH Parallax Results}
\tablehead {
            \colhead{Spot} & \colhead{Offset} &  \colhead{Offset} & \colhead{V$_{LSR}$} &
            \colhead{W3OH/J0230+621} & \colhead{W3OH/J0231+628}&  \colhead{W3OH/J0235+622}
            \\
            \colhead{} & \colhead{East}   & \colhead{North}    &    \colhead{}    &
            \colhead{Parallax} & \colhead{Parallax} & \colhead{Parallax}
            \\

            \colhead{} & \colhead{(mas)}   & \colhead{(mas)} & \colhead{(km/s)}    &
            \colhead{(mas)} & \colhead{(mas)} & \colhead{(mas)}
            }
\startdata
 1  & -16.502 &  -1.317&-43.5 & $0.535 \pm 0.048$ & $0.559 \pm 0.050$  & $0.549 \pm 0.055$ \\
 2  &  -0.145 &   0.163&-43.8 & $0.513 \pm 0.021$ & $0.535 \pm 0.032$  & $0.524 \pm 0.042$ \\
 3  & -11.760 &  -3.456&-44.2 & $0.469 \pm 0.023$ & $0.495 \pm 0.040$  & $0.483 \pm 0.039$ \\
 4  &   0.179 &   0.113&-44.2 & $0.439 \pm 0.018$ & $0.463 \pm 0.037$  & $0.452 \pm 0.036$ \\
 5  &  -7.683 &   8.925&-44.6 & $0.556 \pm 0.044$ & $0.581 \pm 0.052$  & $0.568 \pm 0.045$ \\
 6  &  70.958 &  60.077&-45.8 & $0.493 \pm 0.020$ & $0.515 \pm 0.036$  & $0.505 \pm 0.040$ \\
 7  &  69.766 & -59.651&-42.3 & $0.495 \pm 0.040$ & $0.518 \pm 0.052$  & $0.508 \pm 0.041$ \\
 8  &  69.500 & -59.703&-42.7 & $0.488 \pm 0.024$ & $0.513 \pm 0.041$  & $0.503 \pm 0.045$ \\
 9  &  20.957 &-126.030&-43.1 & $0.529 \pm 0.032$ & $0.552 \pm 0.055$  & $0.543 \pm 0.054$ \\
\enddata

\tablecomments {
Columns 1--4 give the maser spot number, the East and North position offsets
(relative to $\alpha({\rm J2000})=02~27~03.8192$ and
$\delta({\rm J2000})=61~52~25.230$), and the LSR velocity, respectively.
Columns 5--7 give the parallax estimates for W3OH relative to the three extragalactic
sources J0230+621, J0231+628 and J0235+622.
Spot 4 is the reference maser; emission from spots 3--5 can be seen in
Fig.~2, which displays emission in a single spectral channel centered
at $-44.2$~\kms.  Positions and parallaxes for other maser spots were
determined from images in other spectral channels.
}
\end{deluxetable}

\begin{deluxetable}{ccccccccc}
\tablecolumns{9} \tablewidth{0pc} \tablecaption{Proper Motion
Results}
\tablehead{ \colhead{Spot}   & \multicolumn{2}{c}{W3OH/J0230+621} &
                         & \multicolumn{2}{c}{W3OH/J0231+628} &
                         & \multicolumn{2}{c}{W3OH/J0235+622} \\

\colhead{}    & \colhead{$\mu_x$}   & \colhead{$\mu_y$} &
              & \colhead{$\mu_x$}   & \colhead{$\mu_y$} &
              & \colhead{$\mu_x$}   & \colhead{$\mu_y$} \\
\colhead{}    & \colhead{(\masy)}   & \colhead{(\masy)} &
              & \colhead{(\masy)}   & \colhead{(\masy)} &
              & \colhead{(\masy)}   & \colhead{(\masy)} }
              \startdata
1  &$-1.020$ &$-0.134$  &&$-1.125$ &$-0.003$  &&$-1.097$ &$+0.044$  \\
2  &$-1.315$ &$-0.217$  &&$-1.419$ &$-0.085$  &&$-1.392$ &$-0.037$  \\
3  &$-1.068$ &$-0.272$  &&$-1.172$ &$-0.141$  &&$-1.145$ &$-0.093$  \\
4  &$-1.153$ &$-0.340$  &&$-1.257$ &$-0.209$  &&$-1.230$ &$-0.161$  \\
5  &$-1.013$ &$-0.226$  &&$-1.117$ &$-0.095$  &&$-1.090$ &$-0.047$  \\
6  &$-1.112$ &$-0.339$  &&$-1.216$ &$-0.208$  &&$-1.189$ &$-0.160$  \\
7  &$-1.175$ &$-0.257$  &&$-1.279$ &$-0.125$  &&$-1.252$ &$-0.078$  \\
8  &$-1.071$ &$-0.205$  &&$-1.176$ &$-0.074$  &&$-1.148$ &$-0.026$  \\
9  &$-1.363$ &$-0.260$  &&$-1.467$ &$-0.129$  &&$-1.440$ &$-0.081$  \\
\enddata
\tablecomments{
Column 1 gives the maser spot number.  Columns 2 \& 3 give the
motion on the plane of the sky in the eastward and northward directions,
respectively, of a maser relative to the extragalactic source J0230+621.
Columns 3 \& 4 and 5 \& 6 give motions relative to extragalactic sources
J0231+628 and J0235+622.
Typical uncertainties for individual proper motions are $\pm 0.09$ \masy.
The unweighted mean proper motion for W3OH is
$-1.204 \pm 0.02$ \masy\ ($-11.1 \pm 0.2$ \kms) eastward and
$-0.147 \pm 0.01$ \masy\ ($ -1.3 \pm 0.1$ \kms)  northward.
}
\end{deluxetable}

\newpage
{\Large\textbf{Supporting Online Material}}\\
The data were analyzed using the NRAO Astronomical Image
Processing System (AIPS).  The calibration sequence included: 1)
parallactic angle, reference source position, and atmospheric
delay error correction (see below); 2) sampler bias, system
temperature and gain curve corrections; 3) electronic
phase-calibration, phase referencing, and self-calibration.

The main source of systematic error for such observations is
delays introduced to the signals as they propagate through the
Earth's atmosphere and ionosphere {\it [S1]}. Most of the
atmospheric delays are removed when the signals recorded at each
telescope are cross-correlated in the VLBA processor.  However,
the model used in the VLBA processor is a seasonally-averaged
calculation that does not take into account variations in
atmospheric pressure, total water-vapor content, nor the delay
induced by the ionosphere.  At a frequency of 12 GHz, the
processor model will typically be in error by about
0.3~nano-seconds (nsec), corresponding to 10 cm of propagation
path length, for sources at high elevation angles.

Following procedures developed recently {\it [S2]}, we observed
about 15 strong extragalactic radio sources from the International
Celestial Reference Frame catalog {\it [S3]} in rapid succession
over a period of about 45 min at the beginning, middle, and end of
the observing periods. The positions of these sources are known to
better than 1 mas, so as to avoid position errors contributing
significantly to the residual interferometric delays. These data
were taken at 12 GHz with eight 8-MHz bands that spanned 470 MHz
of bandwidth, and residual multi-band delays and fringe rates were
calculated for all sources.  These data were modeled as resulting
from a vertical atmospheric delay and delay-rate, as well as a
clock offset and clock drift rate, at each antenna. We estimated
the zenith atmospheric delay to an accuracy of about 0.01 nsec,
corresponding to 0.3 cm ($\approx0.1\lambda$) of path length.
Since we placed these observing blocks throughout our
observations, we monitored slow changes in the total atmosphere
above each telescope. Correcting for these atmospheric/ionospheric
delays can result in about a factor of five improvement in
relative positions, compared to those obtained by ignoring
atmospheric mis-modeling.

When conducting phase-referenced observations, it is important
that the position of the reference source is well determined.
Since the 12 GHz methanol masers in W3OH are spread over a region
of about 1 arcsec, we needed to determine the absolute position of
the chosen maser spot.  We accomplished this by making a map of
the spectral channel containing the maser spot by simply Fourier
transforming the data following amplitude calibration and
atmospheric model correction, but without phase referencing. This
produces a very crude image, which resembles optical speckles,
owing to the large phase excursions introduced mostly by
short-term fluctuations in the atmospheric/ionospheric path.  For
our data, the brighter speckles were concentrated in a region of
about 30~mas radius, and we were able to estimate the center of
the speckles to an accuracy of about 10 mas. We then shifted the
maser data to remove the position offset of the reference spot
relative to the position used in correlation. After this
correction, the absolute position of the reference phase center
was at $\alpha({\rm J2000})=02~27~03.8192$ and $\delta({\rm
J2000})=61~52~25.230$.

The background sources were selected in two ways.  First, we
searched the on-line VLBA calibrator database and found two
candidates (from the Jodrell Bank--VLA Astrometric Survey),
J0235+622 and J0231+628, which were reasonably strong ($\geq 40$
mJy and 60 mJy, respectively) and near in angle to W3OH ($\sim
1^{\circ}.2$ and $1^{\circ}.5$).  Second, we mapped the region
within $0^\circ.8$ of W3OH with the VLA at wavelengths of 21 and
6~cm in the A-configuration and found a weak compact,
flat-spectrum, source (J0230+621) only $0^{\circ}.7$ from W3OH.

Since our three calibrators had position uncertainties larger than
about 10 mas, we used the absolute position of the maser reference
spot, coupled with our very accurate maser--calibrator offsets, to
determine the absolute position of the calibrators. The absolute
positions of the calibrators determined in this manner are
$\alpha({\rm J2000})=02~30~16.1509$ and $\delta({\rm
J2000})=62~09~37.720$ for J0230+621, $\alpha({\rm
J2000})=02~31~59.1529$ and $\delta({\rm J2000})=62~50~34.227$ for
J0231+628, and $\alpha({\rm J2000})=02~35~20.6356$ and
$\delta({\rm J2000})=62~16~02.350$ for J0235+622.

Images of the three extragalactic sources are shown in Fig.~S1.
All sources are compact ($<0.5$~mas) and dominated by a single
component.

\begin{bf}{References and Notes}\end{bf}\\
S1. M. J. Reid,  A. C. S. Readhead, R. C.  Vermeulen, R. N. Treuhaft, \apj \ 524, 816 (1999)\\
S2. M. J. Reid, A.  Brunthaler, \apj \ 616, 872 (2004)\\
S3. C. Ma,  \etal, \aj \ 116, 516 (1998)\\

\begin{figure}
\figurenum{S1} \epsscale{0.6} \plotone{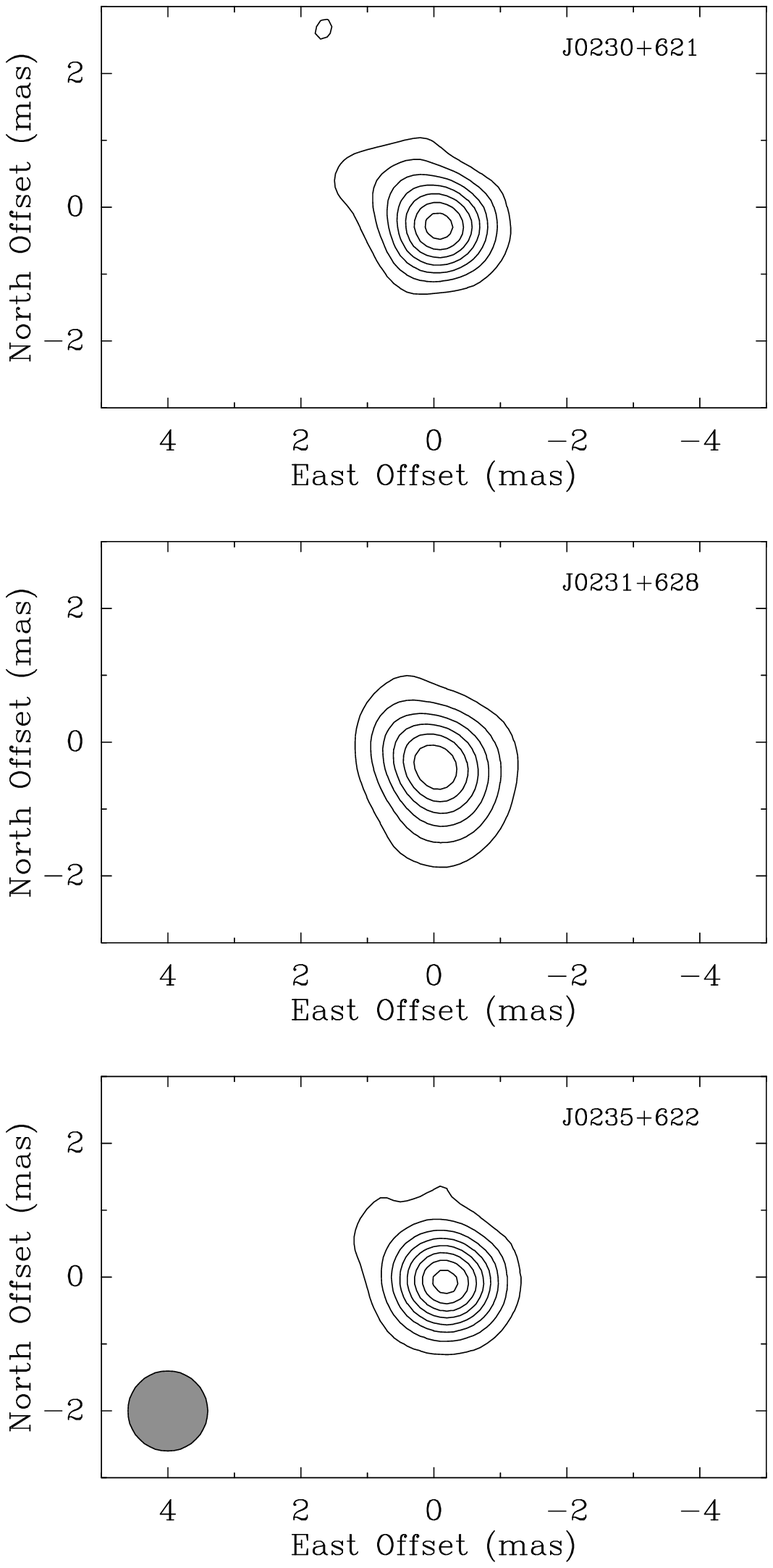} \caption{
Synthesized images of the three background radio sources at the
last epoch. The contour levels are multiples of 2 mJy beam$^{-1}$
for J0230+621 and 5 mJy for J0231+628 and J0235+622, with the zero
contour suppressed. The restoring beam, indicated in the lower
left corner, is a Gaussian with a full-width at half-maximum of
1.2~mas. All sources appear dominated by a single compact
component. \label{S1} }
\end{figure}

\end{document}